\documentclass[twocolumn,prb]{revtex4}

\usepackage[dvips]{graphicx}
\usepackage{amssymb}

\newcommand{\dir}{Figs}

\newcounter{Figure}
\newenvironment{FigureCaptions}{\begin{list}{
  Fig. \theFigure. \rm}
  {\protect\usecounter{Figure}\setlength{\labelwidth}{9em}}
  }{\end{list}}

\newcounter{Table}

\newcommand{\eref}[1]{Eq. \ref{#1}}

\newcommand{\fig}[1]
{
\includegraphics[width=10cm]{fig#1.eps}

\vfill

\noindent
Fig. #1 \\
Cheung and Schmid \\
Journal of Chemical Physics

\clearpage
}

\begin{document}

\newcommand{\CCaa}{
\label{fig:iso_pgs1}
(a) Density profiles for isotropic fluid with $\rho^*=0.24$ and $T^*=0.5$.
Solid line is the simulation data, dashed line shows the data obtained
from DFT calculations using \eref{eqn:excessfe1} and the numerical
DCF, and the dotted line shows the data obtained from DFT calculations 
using \eref{eqn:excessfe2} and the numerical DCF. \\
(b) Order parameter profiles for isotropic fluid with $\rho^*=0.24$ and
$T^*=0.5$. Same symbols as in (a).
}

\newcommand{\CCab}{
\label{fig:iso_pl1}
(a) Density profiles for isotropic fluid with $\rho^*==0.24$ and
$T^*=0.5$. Solid line is the simulation data, dashed line is data
obtained from DFT calculations using \eref{eqn:excessfe1} and the 
Parson-Lee DCF, and the dotted line shows the density profile obtained 
from DFT calculations using \eref{eqn:excessfe2} and the PL DCF. \\
(b) Order parameter profiles for isotropic fluid with $\rho^*=0.24$ 
and $T^*=0.5$. same symbols as in (a).
}

\newcommand{\CCac}{
\label{fig:nem_pgs1}
(a) Density profiles for nematic fluid with $\rho^*=0.30$ and
$T^*=0.5$. Solid line is the simulation data, dashed line is the data
obtained from DFT calculations using \eref{eqn:excessfe1} and
numerical DCF, and the dotted line shows the data obtained from DFT
calculations obtained from DFT calculations using \eref{eqn:excessfe2}
and the numerical DCF. \\
(b) Order parameter profiles for nematic fluid with $\rho^*=0.30$ and
$T^*=0.5$. Same symbols as in (a).
}

\newcommand{\CCad}{
\label{fig:nem_pl1}
(a) Density profiles for nematic fluid. Solid line is the simulation
data ($\rho^*=0.30$ and $T^*=0.5$), dashed line is data obtained from
DFT calculations using \eref{eqn:excessfe1} and Parson-Lee DCF
($\rho^*=0.35$ and $T^*=0.5$), and the dotted line shows the data
obtained from DFT calculations using \eref{eqn:excessfe2} and
Parson-Lee DCF ($\rho^*=0.35$ and $T^*0.5$). \\
(b) Order parameter profiles for nematic fluid. Same symbols as in
(a).
}

\newcommand{\CCae}{
\label{fig:cw1}
(a) Variation of surface free energy with cell width calculated using the
numerical DCF.  The solid line with circles shows the energy of the
isotropic fluid calculated using \eref{eqn:excessfe1}, the dashed line
with circles shows the energy of the isotropic fluid calculated using
\eref{eqn:excessfe2}, the solid line with squares shows the energy of
the nematic fluid calculated using \eref{eqn:excessfe1}, and the
dashed line with squares shows the energy of the nematic fluid
calculated using \eref{eqn:excessfe2}. \\
(b) Variation of the surface free energy with cell width calculated using
the Parsons-Lee DCF. The solid line with circles shows the energy of the
isotropic fluid calculated using \eref{eqn:excessfe1}, the dashed line
with circles shows the energy of the isotropic fluid calculated using
\eref{eqn:excessfe2}, the solid line with squares shows the energy of
the nematic fluid calculated using \eref{eqn:excessfe1}, and the
dashed line with squares shows the energy of the nematic fluid
calculated using \eref{eqn:excessfe2}.
}

\title{
A density functional theory study of the confined soft ellipsoid fluid 
}

\author{David L. Cheung and Friederike Schmid \\
Theoretische Physik, Universit\"{a}t Bielefeld,  \\
33615 Bielefeld, Germany}

\begin{abstract}

A system of soft ellipsoid molecules confined between two planar walls
is studied using classical Density Functional Theory (DFT). Both the
isotropic and nematic phases are considered. The excess free energy is
evaluated using two different {\it Ans\"{a}tze} and the intermolecular
interaction is incorporated using two different direct correlation
functions (DCF). The first is a numerical DCF obtained from
simulations of bulk soft ellipsoid fluids and the second is taken from
Parsons-Lee theory. In both the isotropic
and nematic phases the numerical DCF gives density and order parameter
profiles in reasonable agreement with simulation. The Parsons-Lee DCF
also gives reasonable agreement in the isotropic phase but poor
agreement in the nematic phase.

\end{abstract}

\maketitle


\section{Introduction}

The behaviour of fluids near surfaces and interfaces has attracted 
great interest in recent years. The presence of a surface breaks the 
translational symmetry of the fluid, leading to behaviour radically 
different to that in the bulk. This is particularly true for liquid
crystals \cite{jerome1992a}. Many applications for liquid crystalline
materials rely on the ability to manipulate the preferred alignment
through the action of an external field. This behaviour is strongly 
influenced by coupling between the LC and bounding surfaces.

As may be expected from above, there have been many studies of liquid
crystal systems near surfaces. Experimental studies have been performed
using methods such as scanning tunnelling microscopy
\cite{frommer1992a}, atom force microscopy \cite{williamson1995a} and
NMR \cite{luckhurst1997a} among others. There have also been simulation
studies, using lattice models \cite{zannoni1998a}, hard 
\cite{allen1999a,allen2002a} or soft single site models
\cite{cleaver1997a},  and a few studies using atomistic models 
\cite{cleaver1994a}. It has also been studied by the all
commonly used theoretical methods in liquid crystal science.  Both
elastic theory and Landau-de Gennes \cite{degennes2} theory have been
applied to this. However, the former makes some assumptions
(e.g. slow director variation) that are invalid near a solid
substrate, while the latter introduces phenomenological parameters not
that are not easily related to microscopic properties. Lastly there
have been several studies using density functional theory (DFT)
\cite{sullivan1995a,allen1999a,allen2002a,teixeira1997a,teixeira2001a}
or integral equation methods \cite{lekkerkerker1997a,lekkerkerker1997b}.

In this paper we study a liquid crystal fluid near a solid
substrate using density functional theory (DFT)
\cite{evans1979a,hansenmcdonald}. DFT in principle 
provides a route to the thermodynamic properties of a fluid from
knowledge of its microscopic (molecular) properties. The
intermolecular potential is incorporated through the excess free
energy functional. While the exact form of this is unknown a number of
approximations to this are commonly used, such as Onsager theory
\cite{onsager1949a} and Parsons-Lee theory 
\cite{parsons1979a,lee1987a,lee1988a}.

Previous DFT calculations for liquid crystalline systems have been stymied as
the exact direct correlation function (DCF) is unknown. Commonly this
is approximated by the Mayer $f$-function
\begin{equation}\label{eq:mayerf}
  f(\mathbf{r}_{12},\mathbf{u}_1,\mathbf{u}_2)=
  \exp\left\{
  -\beta V(\mathbf{r}_{12},\mathbf{u}_1,\mathbf{u}_2)\right\}
  -1
\end{equation}
where $V(\mathbf{r}_{12},\mathbf{u}_1,\mathbf{u}_2)$ is the
intermolecular potential and $\beta=\frac{1}{k_BT}$ is the inverse 
temperature. For a hard potential, for which DFT calculations have
mostly been limited to, the integral of the Mayer function gives the excluded
volume between two particles. 

However, recently the DCF has been calculated for the soft
ellipsoid model for a number of state points in both the nematic and
isotropic phase \cite{schmid2001a,schmid2002a,schmid2003a}. It has
been shown to reproduce the bulk properties of the nematic fluid such
as the elastic constants \cite{schmid2001a}. Thus it is hoped that
this may provide a good description of the structure of the
inhomogeneous fluid. 

In this paper we consider one of the simplest situations: a fluid of
ellipsoidal molecules interacting with a structureless
wall. Both isotropic and nematic phases are considered. The paper is
arranged as follows: in the next two sections the theory (Sec. 2) and
computational method are outlined (Sec. 3). The results are then given
in Sec. 4 and a short summary follows in Sec. 5.

\section{Theory}

\subsection{Density Functional Theory}

For a system of uniaxial molecules the Grand Potential can be written as
\begin{eqnarray}
  \label{eqn:grandpot}
  \lefteqn{
  \beta\Omega\left[\rho(\mathbf{r},\mathbf{u})\right]=
  \beta F_{id}\left[\rho(\mathbf{r},\mathbf{u})\right] +
  \beta F_{ex}\left[\rho(\mathbf{r},\mathbf{u})\right] 
  }
  \\ && \nonumber
  + \:\beta \int\;d\mathbf{r} d\mathbf{u}\; 
  V_{ext}(\mathbf{r},\mathbf{u})\rho(\mathbf{r},\mathbf{u})
  - \: \beta\mu\int\;d\mathbf{r} d\mathbf{u}\;
  \rho(\mathbf{r},\mathbf{u})
\end{eqnarray}
where $\rho(\mathbf{r},\mathbf{u})$ is the orientationally dependent
single particle density distribution, $V_{ext}(\mathbf{r},\mathbf{u})$
is the external potential, and $\mu$ is the chemical potential. 
$F_{id}\left[\rho(\mathbf{r},\mathbf{u})\right]$ and
$F_{ex}\left[\rho(\mathbf{r},\mathbf{u})\right]$ are the ideal and
excess free energies, respectively. The ideal free energy is given by
\begin{equation}\label{eqn:idealfe}
  \beta F_{id}\left[\rho(\mathbf{r},\mathbf{u})\right]= 
  \int\;d\mathbf{r}d\mathbf{u}\;
  \rho(\mathbf{r},\mathbf{u})\left[ 
    \log\left(\lambda^3 \rho(\mathbf{r},\mathbf{u})\right)-1
    \right],
\end{equation}
where $\lambda$ is the thermal de Broglie wavelength.
The excess free energy is in general unknown. Here we use two {\it Ans\"{a}tze} for 
$F_{ex}[\rho(\mathbf{r},\mathbf{u})]$. In the first 
$F_{ex}[\rho(\mathbf{r},\mathbf{u})]$ is in the spirit of Onsager
theory \cite{onsager1949a,hansenmcdonald}
\begin{eqnarray}\label{eqn:excessfe1}
\lefteqn{
  \beta F_{ex}\left[\rho(\mathbf{r},\mathbf{u})\right]=
} \\ &&
  -\frac{1}{2}
  \int\;d\mathbf{r}_1d\mathbf{u}_1d\mathbf{r}_2d\mathbf{u}_2 \;
  c(\mathbf{r}_{12},\mathbf{u}_1,\mathbf{u}_2)
  \rho(\mathbf{r}_1,\mathbf{u}_1)\rho(\mathbf{r}_2,\mathbf{u}_2),
  \nonumber
\end{eqnarray}
where $c(\mathbf{r}_{12},\mathbf{u}_1,\mathbf{u}_2)$ is the direct
correlation function.

In the second $F_{ex}[\rho(\mathbf{r},\mathbf{u})]$ is taken to be a
density expansion around the homogenous fluid with density
$\rho_0(\mathbf{u})$ truncated at the 2nd order term \cite{sweatman2000a}
\begin{eqnarray}
\nonumber
 \lefteqn{
  \beta F_{ex}\left[\rho(\mathbf{r},\mathbf{u})\right]=
  -\int\;d\mathbf{r}_1d\mathbf{u}_1
  c^{(1)}(\mathbf{u}_1)
  \left[\rho(\mathbf{r}_1,\mathbf{u}_1)-\rho_0(\mathbf{u})\right]
}\qquad \\ 
  \label{eqn:excessfe2}
  &-&
  \frac{1}{2}
  \int\;d\mathbf{r}_1d\mathbf{u}_1d\mathbf{r}_2d\mathbf{u}_2 \; 
  c(\mathbf{r}_{12},\mathbf{u}_1,\mathbf{u}_2) 
  \\ &&
  \times
  \left\{\rho(\mathbf{r}_1,\mathbf{u}_1)-\rho_0(\mathbf{u}_1)\right\}
  \left\{\rho(\mathbf{r}_2,\mathbf{u}_2)-\rho_0(\mathbf{u}_2)\right\},
  \nonumber
\end{eqnarray}
where $\rho_0(\mathbf{u})$ is the density for the bulk, homogenous
fluid and $c^{(1)}(\mathbf{u})$ is the first order direct correlation
function. This can be identified with the excess chemical potential (
$\beta\mu_{ex}=-c^{(1)}$) \cite{sweatman2000a}.

For fixed external and chemical potentials, the equilibrium single
particle density is that which minimises the Grand Potential. This is
a solution to the Euler-Lagrange equation
\begin{equation}\label{eqn:euler1}
  \frac{\delta\Omega\left[\rho(\mathbf{r},\mathbf{u})\right]}
       {\delta\rho(\mathbf{r},\mathbf{u})}
  =0
\end{equation}
\begin{equation}\label{eqn:euler2}
  \log\left\{\lambda^3\rho(\mathbf{r},\mathbf{u})\right\}-
  \beta\mu(\mathbf{u})+
  \beta V(\mathbf{r},\mathbf{u})+
  \frac
  {\delta\left\{\beta F_{ex}\left[\rho(\mathbf{r},\mathbf{u})\right]\right\}}
  {\delta \rho(\mathbf{r},\mathbf{u})}=0
\end{equation}

The chemical potential is found from the Euler-Lagrange equation of
the bulk, homogenous fluid, i.e. substituting
$\rho(\mathbf{r},\mathbf{u})=\rho_0(\mathbf{u})$ into
\eref{eqn:euler2} for both {\it Ans\"{a}tze} for the excess free energy. 

The external potential is given by a repulsive 
Lennard-Jones potential
\begin{equation}
  V_{ext}(z) =  
  \left\{
  \begin{array}{c @{ \:\:} c} 
    4\left[
      \left(\frac{1}{z+0.5}\right)^{12}-
      \left(\frac{1}{z+0.5}\right)^6
      \right]
    \;, \mbox{if $z<2^{\frac{1}{6}}-0.5$} \\
    0\:, \mbox{otherwise}
  \end{array}
  \right.
\end{equation}
This potential acts upon the molecular centres, thus giving rise to
homeotropic (normal) alignment at the surface.

\subsection{Direct Correlation Function}

The direct correlation function (DCF),
$c(\mathbf{r}_{ij},\mathbf{u}_i,\mathbf{u}_j)$, is the central
quantity in many theories of liquids and liquid crystals. For liquid
crystalline systems
the simplest approximation to the DCF is that of Onsager where the DCF
is taken to be the Mayer $f$-function. This corresponds to truncating
the virial expansion at the second term, and thus is valid only for
low densities and large elongations.

In this paper, two different forms of the DCF are considered. The
first is a numerical DCF calculated from simulations of soft ellipsoid
molecules \cite{schmid2001a,schmid2002a,schmid2003a}. In principle
this should be an exact representation of the interactions within the fluid. 

The second form is taken from Parsons-Lee theory \cite{parsons1979a, lee1987a,lee1988a}. This corresponds to
a resummed virial series truncated, as in Onsager theory, at the second
order term. Thus the Parsons-Lee DCF is the Mayer $f$-function multiplied
be a density dependent prefactor
\begin{equation}
  c^{PL}(\mathbf{r}_{12},\mathbf{u}_1,\mathbf{u}_2)=
  \frac{\eta(4-3\eta)}{(1-\eta)^2}
  f(\mathbf{r}_{12},\mathbf{u}_1,\mathbf{u}_2),
\end{equation}
where $\eta=\rho v_{mol}$ is the packing fraction.
To calculate the Mayer function the intermolecular potential is
required. For the soft ellipsoids considered in this paper, the
potential is given by
\begin{eqnarray}
  V(\mathbf{r}_{12},\mathbf{u}_1,\mathbf{u}_2)&=&
  4\left[
    \left(
    \frac{\sigma_0}
    {r_{12}-\sigma(\mathbf{\hat{r}_{12}},\mathbf{u}_1,\mathbf{u}_2)+\sigma_0}
    \right)^{12}- \right. \nonumber\\
    & &
    \left.
    \left(
    \frac{\sigma_0}
    {r_{ij}-\sigma(\mathbf{\hat{r}_{12}},\mathbf{u}_1,\mathbf{u}_2)+\sigma_0}
    \right)^6
    \right],
\end{eqnarray}   
where
\begin{eqnarray}
\lefteqn{
  \sigma(\mathbf{\hat{r}}_{ij},\mathbf{u}_i,\mathbf{u}_j)=
  \sigma_0
} \\ \nonumber
  &\times& \!\!
  \left[
    1-\frac{\chi}{2}\left\{
    \frac{
      (\mathbf{\hat{r}}_{ij}.\mathbf{u}_i+\mathbf{\hat{r}}_{ij}.\mathbf{u}_j)^2
    }
    {1+\chi\mathbf{u}_i.\mathbf{u}_j}
    +
    \frac{
      (\mathbf{\hat{r}}_{ij}.\mathbf{u}_i-\mathbf{\hat{r}}_{ij}.\mathbf{u}_j)^2
    }
    {1-\chi\mathbf{u}_i.\mathbf{u}_j}
    \right\}
    \right]^{-\frac{1}{2}},
\end{eqnarray}
and $\chi=(\kappa^2-1)/(\kappa^2+1)$, where $\kappa$ is molecular length to
breadth ratio, equal to 3 for the system considered here.

In both theories the DCF is taken to be that of the bulk, homogenous
fluid (i.e. the density in the prefactor in Eq. 9 is fixed to be the
bulk density). Extension of Parsons-Lee theory to inhomogenous
densities (i.e. a prefactor that depends on the spatially varying
density) is possible although non-trivial
\cite{somoza1989a,loewen1999a,velasco2003a} and application of this is
beyond the scope of this paper.

\section{Method}

Following Allen \cite{allen1999a,allen2000a,allen2002a}, angular dependent functions are expended in spherical harmonics. The
single particle density and its logarithm become
\begin{equation}\label{eqn:rhosp1}
  \rho(\mathbf{r},\mathbf{u}) =
  \sum_{l,m} \rho_{lm}(\mathbf{r})Y_{lm}^*(\mathbf{u})
\end{equation}
and
\begin{equation}\label{eqn:lrhosp1}
  \log\left\{\lambda^3\rho(\mathbf{r},\mathbf{u})\right\} =
  \sum_{l,m} \tilde{\rho}_{lm}(\mathbf{r})Y_{lm}(\mathbf{u}).
\end{equation}

Similarly, the direct correlation function is expanded as
\begin{eqnarray}\label{eqn:dcfsp1}
\lefteqn{
  c(\mathbf{r}_{12},\mathbf{u}_1,\mathbf{u}_2) = 
} \\ && \nonumber
  \sum_{l_1,l_2,l,m_1,m_2,m} c_{l_1m_1l_2m_2lm}(r_{12})
  Y_{l_1m_1}(\mathbf{u}_1)Y_{l_2m_2}(\mathbf{u}_2)Y_{lm}(\mathbf{\hat{r}}_{12})
  .
\end{eqnarray}
In the above sums, $0 \leq l \leq l_{max}$ and $-l \leq m \leq l$.
Symmetry dictates that 
$m_1+m_2+m=0$ and that $l_1$, $l_2$, and $l$ are even.

For a homogenous fluid confined between two structureless walls with
the wall normal in the $z$ direction the density and its logarithm
are functions of $z$ only. 

For the first {\it Ansatz} to the excess free energy, substituting \eref{eqn:rhosp1}, 
\eref{eqn:lrhosp1}, and \eref{eqn:dcfsp1} into the grand potential 
\eref{eqn:grandpot} and performing angular and $x$ and $y$
integrations gives
\begin{widetext}
\begin{eqnarray}
\nonumber
  \frac{\beta\Omega\left[\rho(z,\mathbf{u})\right]}{A}&=&
  \int_0^L\;dz\; \sum_{l,m} 
  \rho_{lm}(z)\left\{\tilde{\rho}_{lm}(z)-\sqrt{4\pi}\delta_{l0}\right\}
  + 
  \sum_{l,m} \beta \int_0^L\;dz\; V_{lm}(z)\rho_{lm}(z)-
  \sqrt{4\pi}\beta\mu\int_0^L\;dz\;\rho_{lm}(z) \\
  && - \sum_{l_1,l_2,m}\frac{1}{2}\int_0^L\;dz_1\;\int_0^L\;dz_2\;
  C_{l_1l_2m}(|z_1-z_2|)\rho_{l_1m}(z_1)\rho_{l_2\bar{m}}(z_2).
  \label{eqn:grandpot2}
\end{eqnarray}
\end{widetext}
The $V_{lm}(z)$ are the components of the spherical harmonics
expansion of the external potential (
$V(z,\mathbf{u})=\sum_{l,m} V_{lm}(z)Y_{lm}(\mathbf{u})$). For the
potential given above, the only nonzero term is 
$V_{00}(z)=\sqrt{4\pi}V_{ext}(z)$. 

$C_{l_1l_2m}(z)$ is the integral of the DCF over $r$ and $\phi$ for a
fixed $z$, given by \\

\noindent
\parbox{0.51\textwidth}{
\begin{equation}
  C_{l_1l_2m}(z)=\sum_{l}\!\!\sqrt{(2l+1)\pi}\int_z^\infty \!\!\!\!\! dr\; 
  rc_{l_1ml_2\bar{m}l0}(r) P_l\left(\frac{z}{r}\right)
\end{equation}
}
where $P_l(x)$  is the $l$th Legendre polynomial.

Similarly the Grand Potential for the second {\it Ansatz} becomes
\begin{widetext}
\begin{eqnarray}
\nonumber
  \frac{\beta\Omega\left[\rho(z,\mathbf{u})\right]}{A} &=&
  \int_0^L\;dz\; \sum_{l,m}
  \rho_{lm}\left\{\tilde{\rho}_{lm}(z)-\sqrt{4\pi}\delta_{l0}\right\}
  +\sum_{l,m} \beta \int_0^L\;dz\;V_{lm}(z)\rho_{lm}(z) - 
  \sum_{l,m} \beta \mu_{lm}\int_0^L\;dz\; \rho_{lm}(z) 
  \\ && \nonumber
  -\frac{1}{2}\sum_{l_1,l_2,m}
  \int_0^L\;dz_1\;\int_{-\infty}^\infty\;dz_2\;
  C_{l_1l_2m}(|z_1-z_2|)\rho_{l_1m}(z_1)
  \left[\rho_{l_2\bar{m}}(z_2)-2\rho^0_{lm}\right]
  \\&&
  - \frac{1}{2}\sum_{l_1,l_2,m}\int_0^L dz_1\;\int_{-\infty}^\infty dz_2
  C_{l_1l_2m} \rho_{l_1m}^0\rho_{l_2\bar{m}}
  \label{eqn:grandpot3}
\end{eqnarray}
\end{widetext}
where $\mu_{lm}= \tilde{f}_{l}\delta_{m0}$ and
$\tilde{f}_l$ are the spherical harmonics coefficients of the
logarithm of the bulk ODF. Note that here $\mu$ includes contriubtions
from the linear term in the excess free energy \eref{eqn:excessfe2}
and thus depends on orientation.

The equilibrium particle densities are found by tabulating the grand
potential on a regular grid and then numerically minimising 
\eref{eqn:grandpot2} or \eref{eqn:grandpot3} with respect to the 
$\tilde{\rho}_{lm}(z)$. This numerical minimisation is performed using 
the conjugate gradients method \cite{numericalrecipies}. The
calculations were taken to be converged when the fractional change in energy
between iterations was less than 0.001 (i.e. $E_{new}-E_{old} < 0.001$). Spatial integrals where
evaluated using the trapezium rule with a grid spacing of 0.04, while
angular integrations were performed using Gauss-Legendre quadrature
with 800 to 6400 points.

Once the equilibrium orientationally dependent density
$\rho(z,\mathbf{u})$ is known, the bulk density can be calculated by
integration over the angular variables, i.e.
\begin{eqnarray}
  \rho_0(z)&=&\int\; d\mathbf{u}\;\rho(z,\mathbf{u}) \nonumber\\
           &=&\sqrt{4\pi}\rho_{00}(z).
\end{eqnarray}
Similarly the 2nd rank order parameter profile, $p_2(z)$, can be 
calculated from \cite{teixeira2001a}
\begin{eqnarray}
  p_2(z)&=& \frac
  {\int\; d\mathbf{u}\rho(z,\mathbf{u})P_2(\mathbf{u})}
  {\int\; d\mathbf{u}\rho(z,\mathbf{u})} 
  \nonumber\\
  p_2(z)&=&\frac{1}{\sqrt{5}}\frac{\rho_{20}(z)}{\rho_{00}(z)}.
\end{eqnarray}

\section{Results}

\subsection{Density and Order Parameter Profiles}

\subsubsection{Isotropic Phase}

\begin{figure}[t]
\hspace*{6mm}\includegraphics[width=6cm]{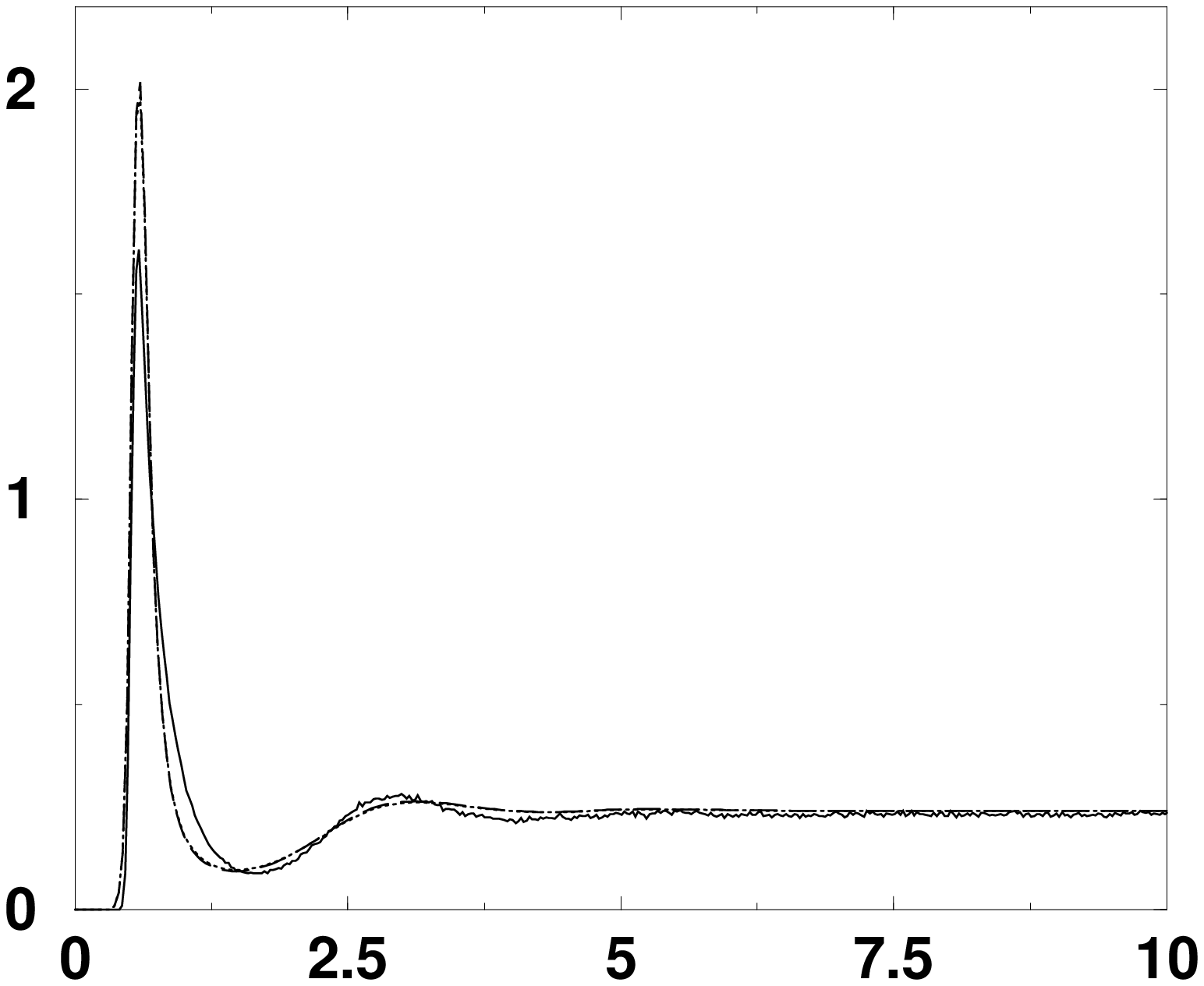}
\vspace*{5mm}
\includegraphics[width=6.5cm]{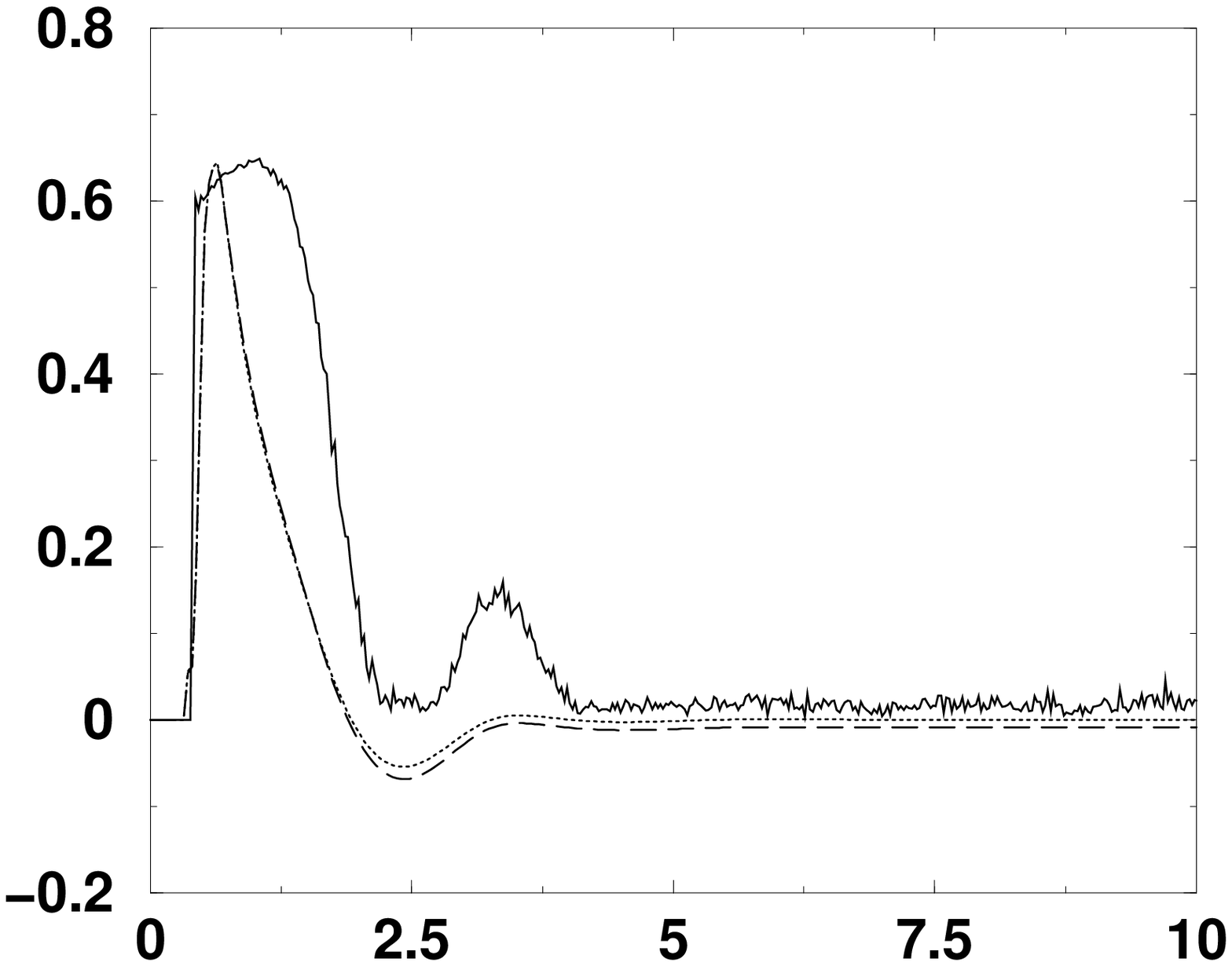}
\caption{ \CCaa }
\end{figure}

First we present results for calculations performed in the isotropic 
phase, with $\rho^*=0.24$ and $T^*=0.5$. Shown in
Fig.~\ref{fig:iso_pgs1}  are the density and order parameter profiles obtained from
DFT calculations. Shown for comparison are
profiles found from molecular dynamics simulations of a confined
system of soft ellipsoids. 

As can be seen from Fig. \ref{fig:iso_pgs1}a the density profiles
found from DFT calculations using \eref{eqn:excessfe1} and
\eref{eqn:excessfe2} are virtually indistinguishable. The DFT profiles
share the same gross features as those 
obtained from simulation. There is a
strong maximum at about $z=0.68$, followed by a minimum at about
$z=1.44$. There is also a weak maximum at about $z=3.4$ in the DFT
profiles and $z=3$ in the simulation data. Both the DFT and simulation
profiles rapidly tend towards the bulk density. 

The most obvious difference between the DFT and simulation results is the
height of the first peak. In the profile obtained from the DFT
calculations the height of the first peak is about 2.02, while in 
the simulation data the it is approximately 1.61.

Shown in Fig. \ref{fig:iso_pgs1}b are the order parameter profiles
obtained from DFT calculations and simulation. As can be seen the
agreement 
between the two profiles here
is not as good as for the density. Both the DFT and simulation data
have peaks at about $z=0.64$. This peak arises due to the effect of
the surface on the ordering of the molecules. The heights of these
peaks are both about $p_2=0.64$. This initial peak is then followed by 
a decrease in the order parameter.

In the DFT calculations the order parameter then
becomes negative with a minimum at about $z=2.4$. The profile found
from simulation has a minimum at about the same distance, although in
this case the order parameter remains positive. Another maximum is
present at about $z=3.3$ in the simulation data. this is also present
in the profiles obtained from DFT calculations, although of much
smaller magnitude. Both profiles then rapidly decay to a constant value
close to 0 indicating isotropic ordering in the bulk of the cell. 

{
\begin{figure}
\hspace*{2mm} \includegraphics[width=6cm]{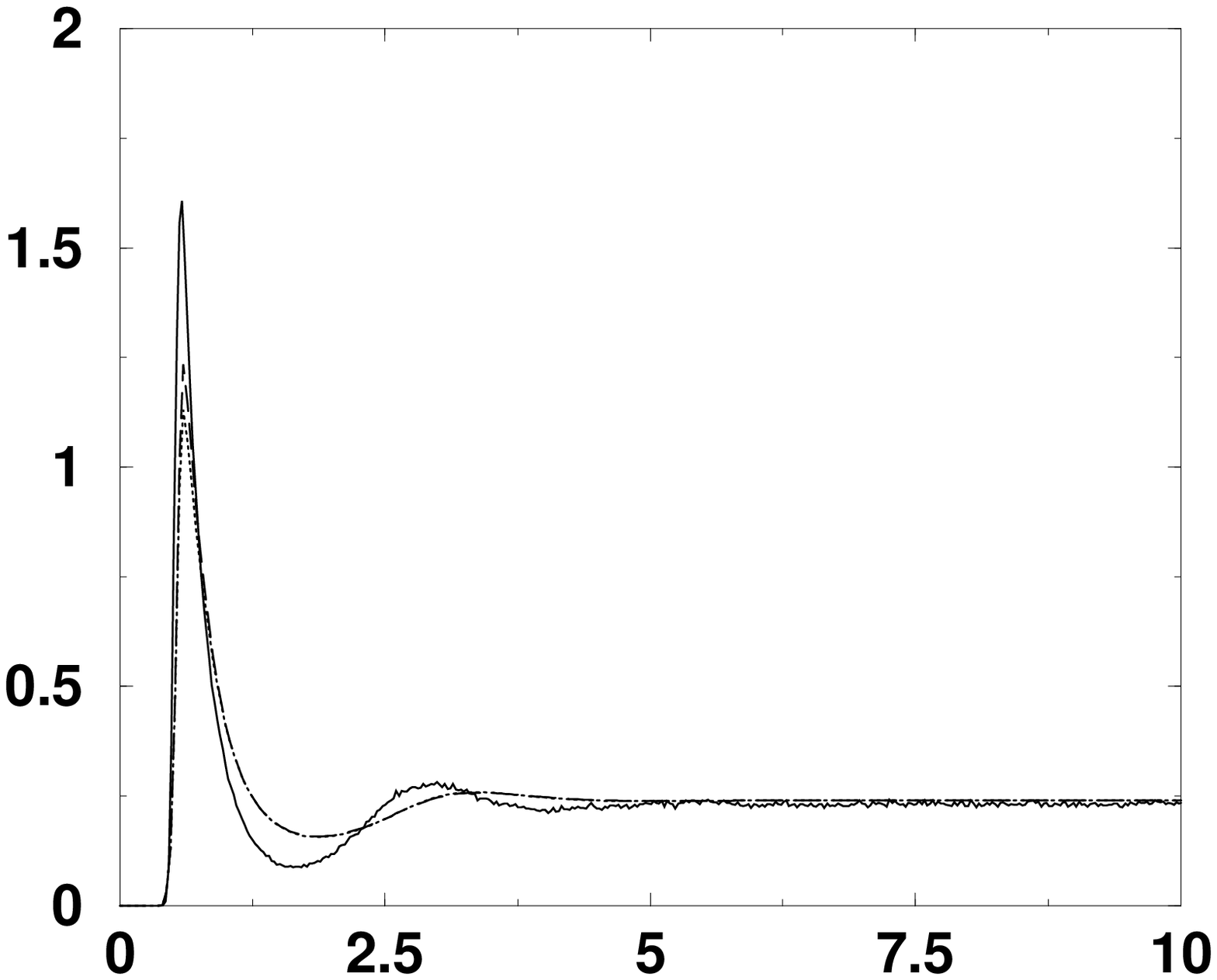}
\vspace*{5mm}
\includegraphics[width=6.15cm]{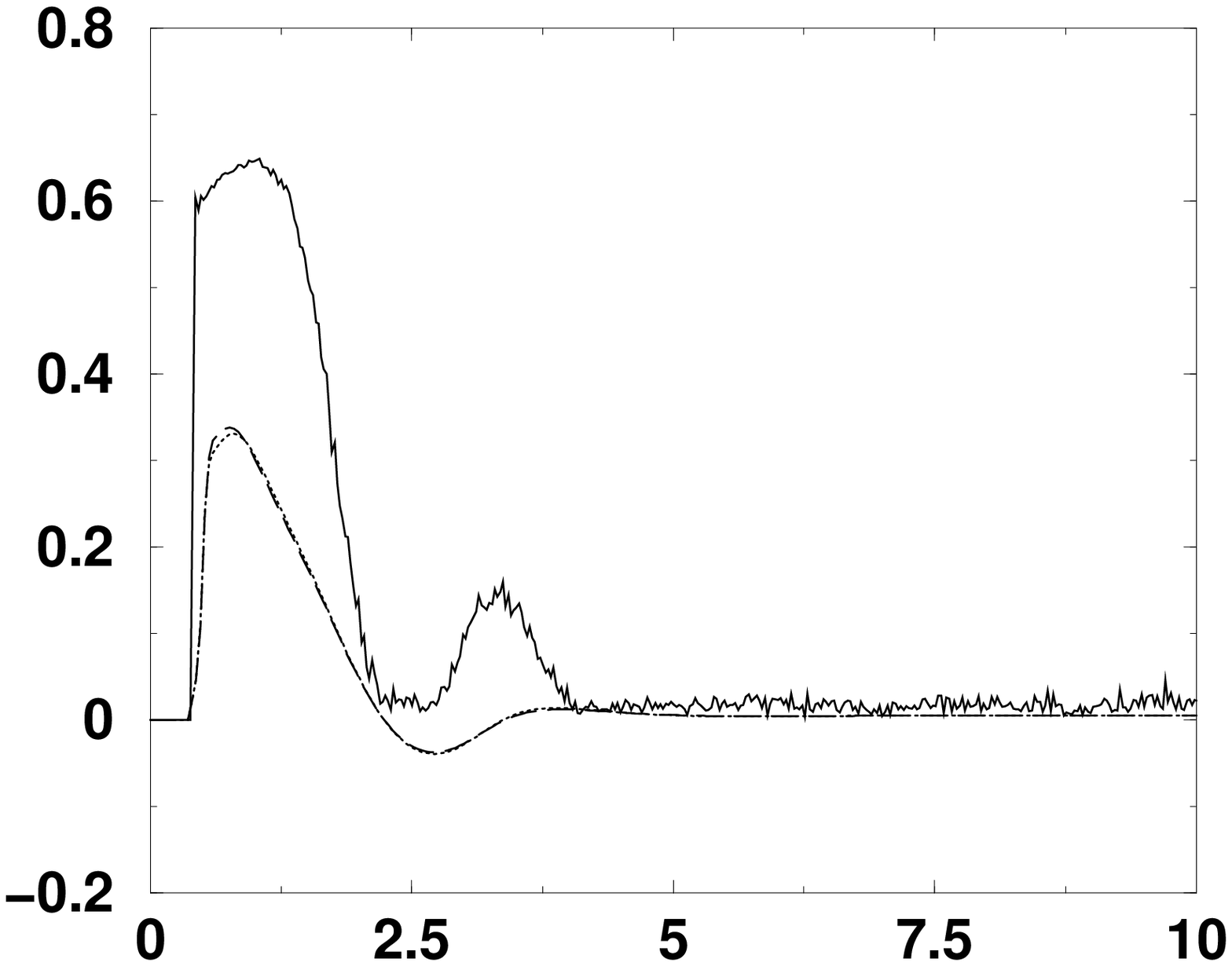}
\caption{ \CCab }
\end{figure}
}

Shown in Fig. \ref{fig:iso_pl1} are the density and order parameter
profiles obtained from DFT calculations using the Parson-Lee DCF. As
for those obtained using the numerical DCF, the density profiles
(Fig. \ref{fig:iso_pl1}a) using Eqs. \ref{eqn:excessfe1} 
and \ref{eqn:excessfe2} are essentially identical and in good
agreement with the profile obtained from simulation. Again a major
discrepancy between the DFT and simulation profiles is the height of
the first peak. In this case however, the peak height is
underestimated by the DFT calculations, 1.23 compared to 1.61 for the
simulation profile.

The order parameter profiles are shown in Fig. \ref{fig:iso_pl1}. As
can be seen the first peak in the order parameter profiles obtained
from DFT calculations are much smaller than in the simulation profile
and those found from DFT calculations using the numerical DCF. For
molecules with length to width ration of 3, Parsons-Lee theory predicts
an isotropic-nematic transition at about $\rho = 0.32$ \cite{lee1988a}, whereas for
simulation the transition is at $\rho = 0.284$. Thus for 
Parsons-Lee theory it is further from the nematic phase, and hence at
lower order, than in simulation.

\subsubsection{Nematic Phase}

{
\begin{figure}
\hspace*{5mm}\includegraphics[width=6cm]{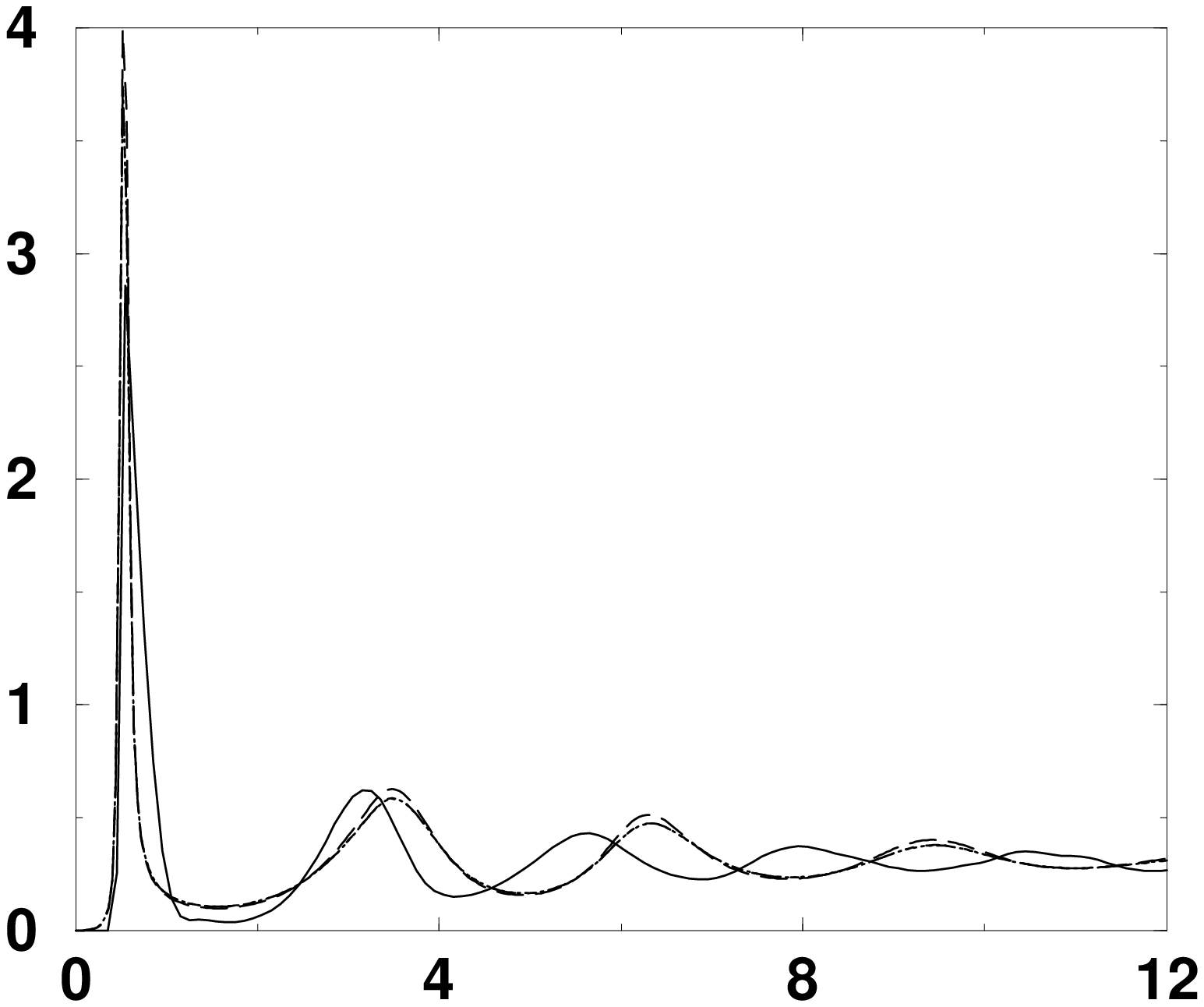}
\vspace*{5mm}
\includegraphics[width=6.5cm]{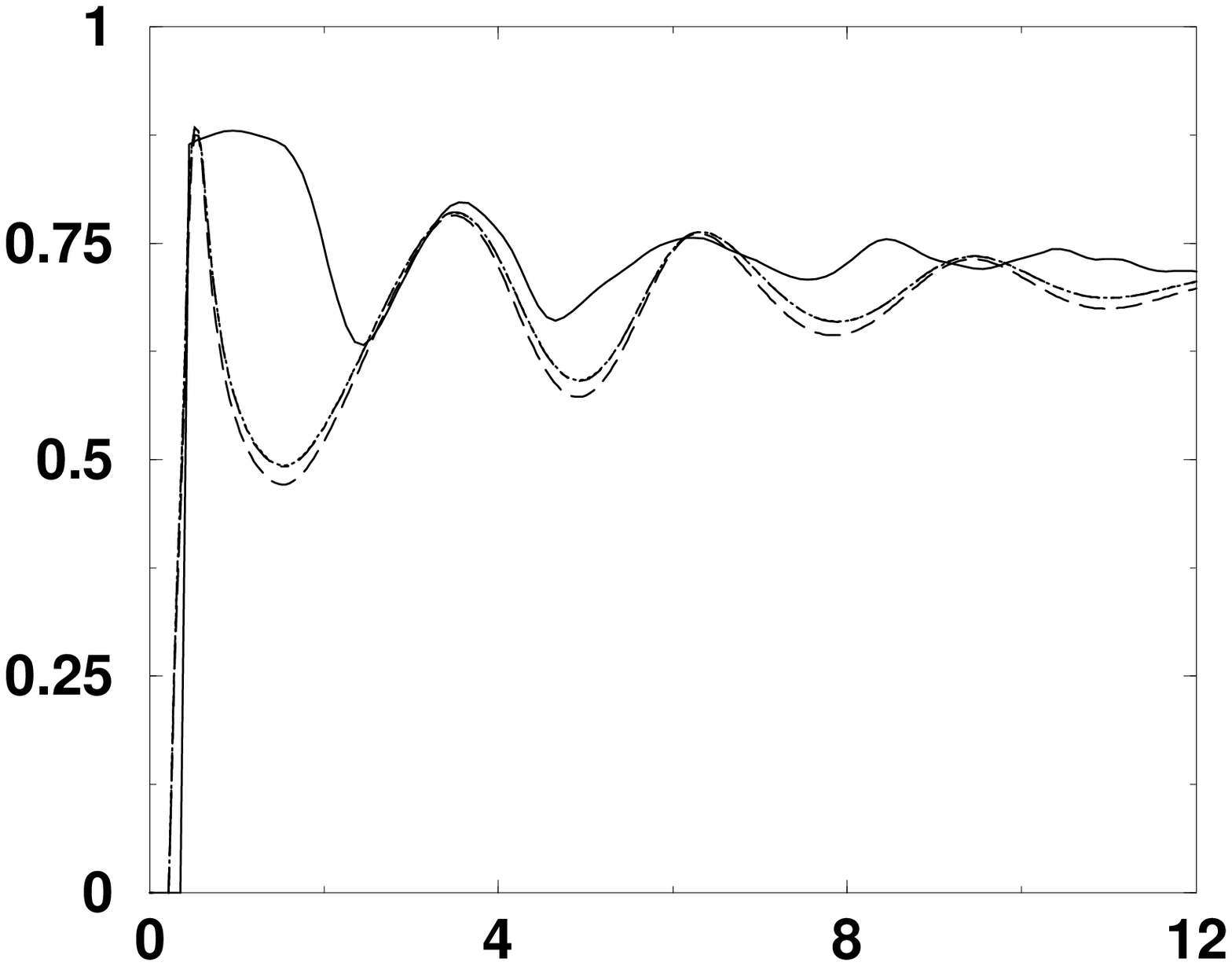}
\caption{ \CCac }
\end{figure}
}

Shown in Fig. \ref{fig:nem_pgs1} are the density and order parameter 
profiles for the nematic phase ($\rho^*=0.30$ and $T^*=0.5$)
calculated using found from DFT calculations and Monte Carlo
simulation. In the simulations, the cell seize was adjusted so that
the average density in the cell bulk was 0.30 to aid comparison
between simulation and theory. As can be
seen the profiles obtained from DFT calculations using
Eqs. \ref{eqn:excessfe1} and \ref{eqn:excessfe2} are in better agreement
with each other than with the simulation profiles.

The density profiles are shown  Fig. \ref{fig:nem_pgs1}a. As in the
isotropic case, there is an initial peak in the density
profile at about $z = 0.6$. There are then further peaks. In the DFT 
profiles these subsequent peaks are at $z = 3.5$, $z = 6.3$, and
$z = 9.4$. In the simulation profile, these are at
$z = 3.2$, $z = 5.6$, and $z = 7.9$. A similar difference has been
seen in comparison between DFT calculations and simulations for simple
fluids\cite{tarazona1985a,vanderlick1989a}. In \cite{tarazona1985a}
this was explained due to the imposition of homogeneity in the
transverse ($x$ and $y$) directions in the theoretical calculations.

Aside from the positions of these secondary peaks, the other noticeable
difference between the DFT and simulation profiles are in the heights
of these peaks. As in the isotropic case the height of the initial
peak is overestimated compared to simulation. The initial peak in the 
DFT calculations has a height of 3.9, while in the 
simulation profile its height is 2.9.  In contrast to the isotropic
phase, the 
secondary
peaks in the DFT profile are more noticeable than those in the
simulation profile. These secondary peaks are also stronger in the DFT
profile calculated using \eref{eqn:excessfe1} than those in the
profile calculated using \eref{eqn:excessfe2}.
These peaks indicate surface induced layering \cite{jerome1992a}. This
has been observed in by X-ray reflectivity measurements, both at the
nematic-solid interface as studied here and at the free nematic interface.

The second rank order parameter profiles are shown in 
Fig. \ref{fig:nem_pgs1}b . Again these show a strong initial peak at
approximately $z=0.6$. This is then follow by a series of peaks before
decaying to a bulk values as for the density profiles. These peaks are
at approximately the same positions as the peaks in the density
profiles. As for the density profile the secondary peaks are stronger
than those in the simulation profile. In the centre of the cell the
DFT order parameters reach constant values. For \eref{eqn:excessfe1}
the order parameter in the centre of the cell is 0.65, whereas for
\eref{eqn:excessfe2} $p_2(z)=0.69$. These values are close to the bulk
of 0.69. The order parameter in the centre of the cell from simulation
is 0.74, higher than the bulk value.

One major difference between the DFT and simulation profiles is the 
decay of the initial peak. In the DFT profile there is a rapid decay
from the first peak that is not seen in the simulation profile. This
can also be seen, albeit not as dramatically, in the order parameter
profiles for the isotropic fluid (Figs. \ref{fig:iso_pgs1} and 
\ref{fig:iso_pl1}). 

{
\begin{figure}
\hspace*{6mm}\includegraphics[width=6cm]{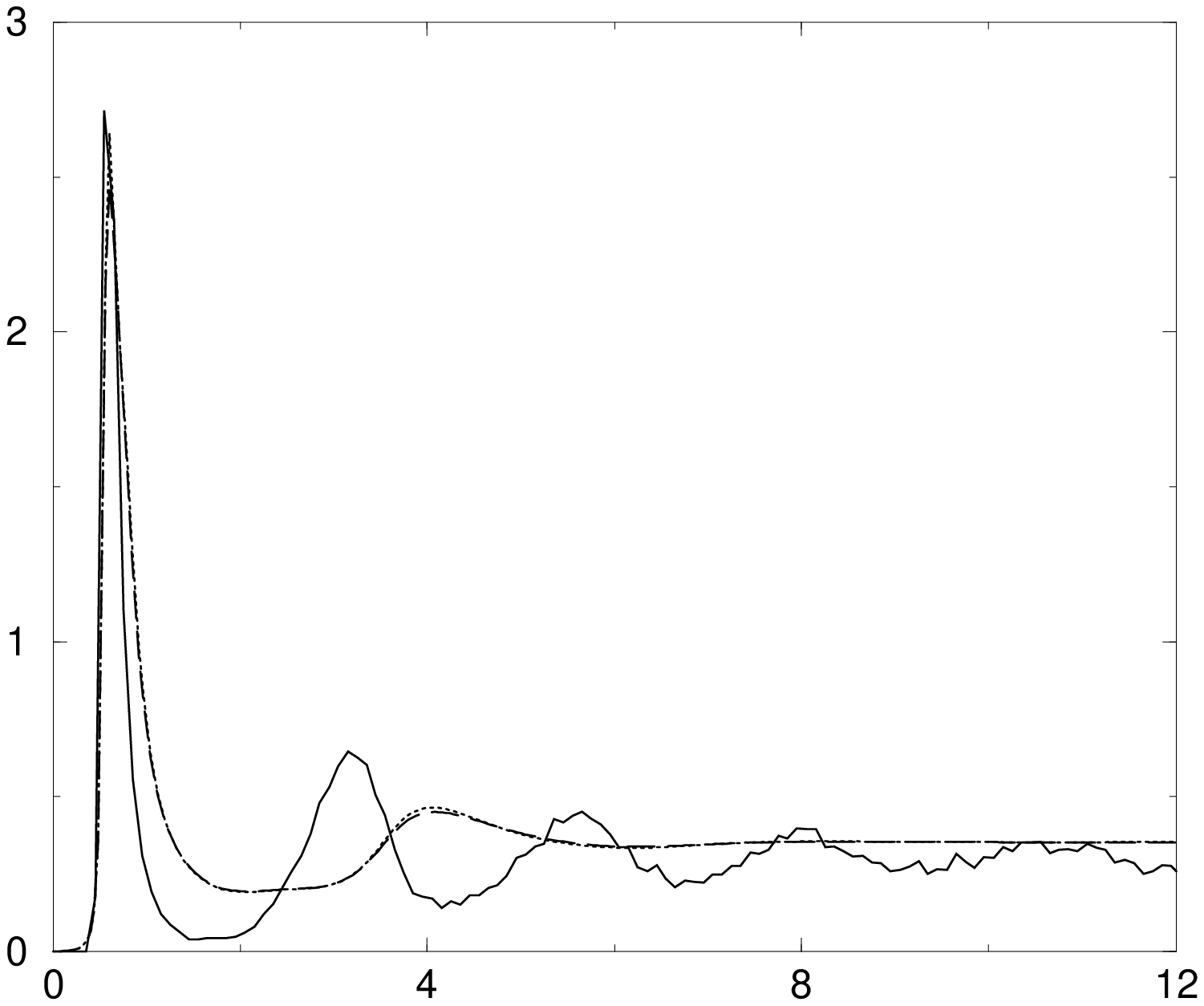}
\vspace*{5mm}
\includegraphics[width=6.6cm]{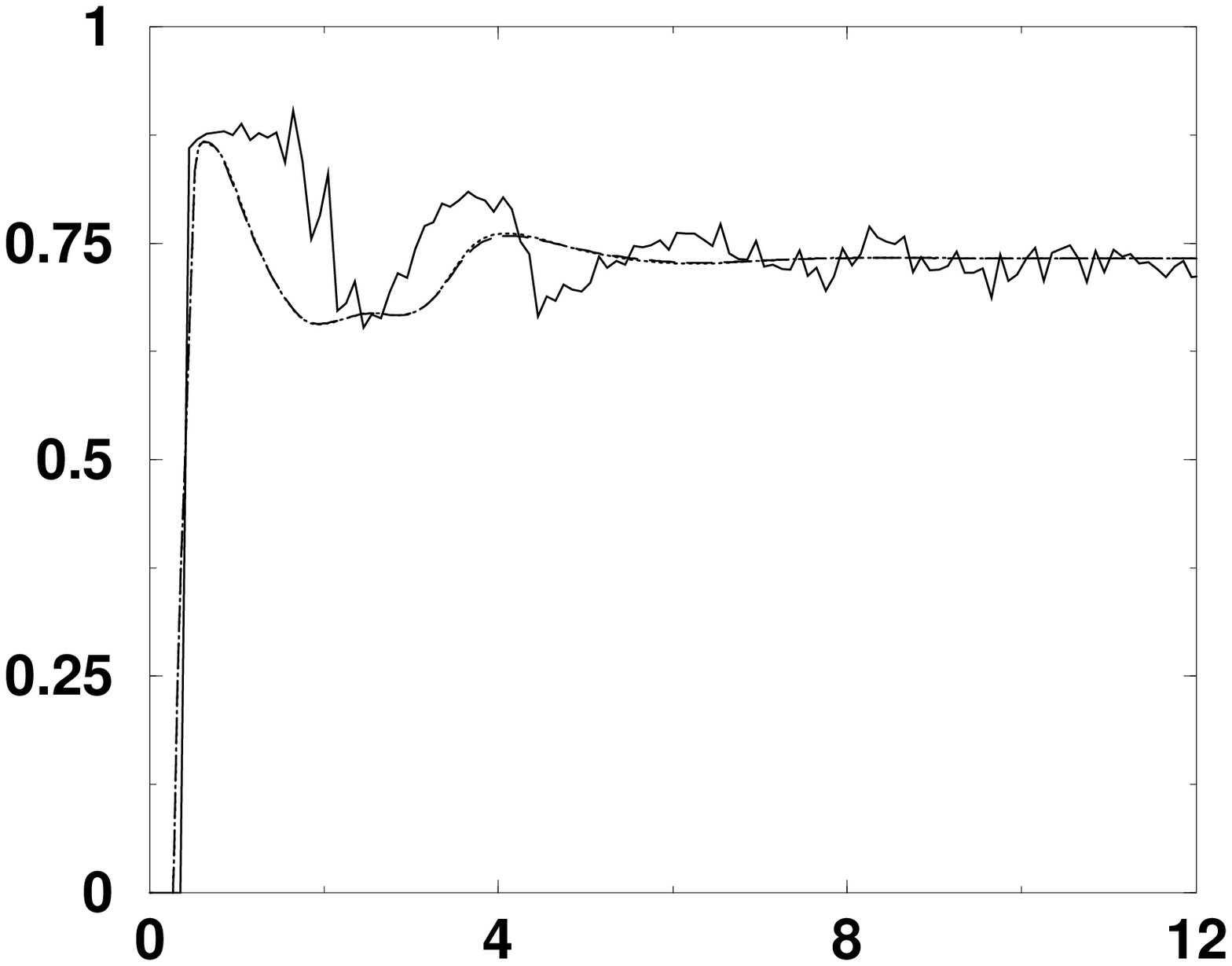}
\caption{ \CCad }
\end{figure}
}

Shown in Fig. \ref{fig:nem_pl1} are the density and order parameter
profiles calculated using the Parson-Lee DCF and those from simulation.
For this model, Parson-Lee theory predicts that a system with 
$\rho^*=0.30$ is isotropic \cite{lee1988a}, these calculations were 
performed with a higher bulk density ($\rho^*=0.35$). While this
difference makes detailed comparison impossible, it is small enough for
qualitative comparison.

As can be seen the density and order parameter profiles calculated
using Eqs. \ref{eqn:excessfe1} and \ref{eqn:excessfe2} are essentially 
identical and are similar to the profiles calculated in the isotropic
phase using the Parsons-Lee DCF. This can be understood, as the
Parsons-Lee DCF is proportional to the Mayer function at all
densities, so  aside from a constant the Parsons-Lee DCF is the same
in both the isotropic and nematic
phases. This leads to a lack of structure in the nematic density and order
parameter profiles compared to those from simulation and DCF
calculations using the numerical DCF. 

\subsection{Variable Cell Width}

In this section, the effect of varying the width of the cell is examined.
A fluid confined between two planar walls is often used in the
evaluation of the depletion force between two colloidal particles
\cite{lekkerkerker1997a}. Here
the surface free energy $\gamma$, given by the excess (over bulk) grand
potential per unit area \cite{hansenmcdonald}
\begin{equation}
  \beta \gamma =\frac
        {\beta\Omega[\rho(\mathbf{z},\mathbf{u})]-
         \beta\Omega[\rho_0(\mathbf{u})]}
        {2A},
\end{equation}
is calculated for cell widths from 4 to 40 $\sigma$. 

{
\begin{figure}
\includegraphics[width=6.2cm]{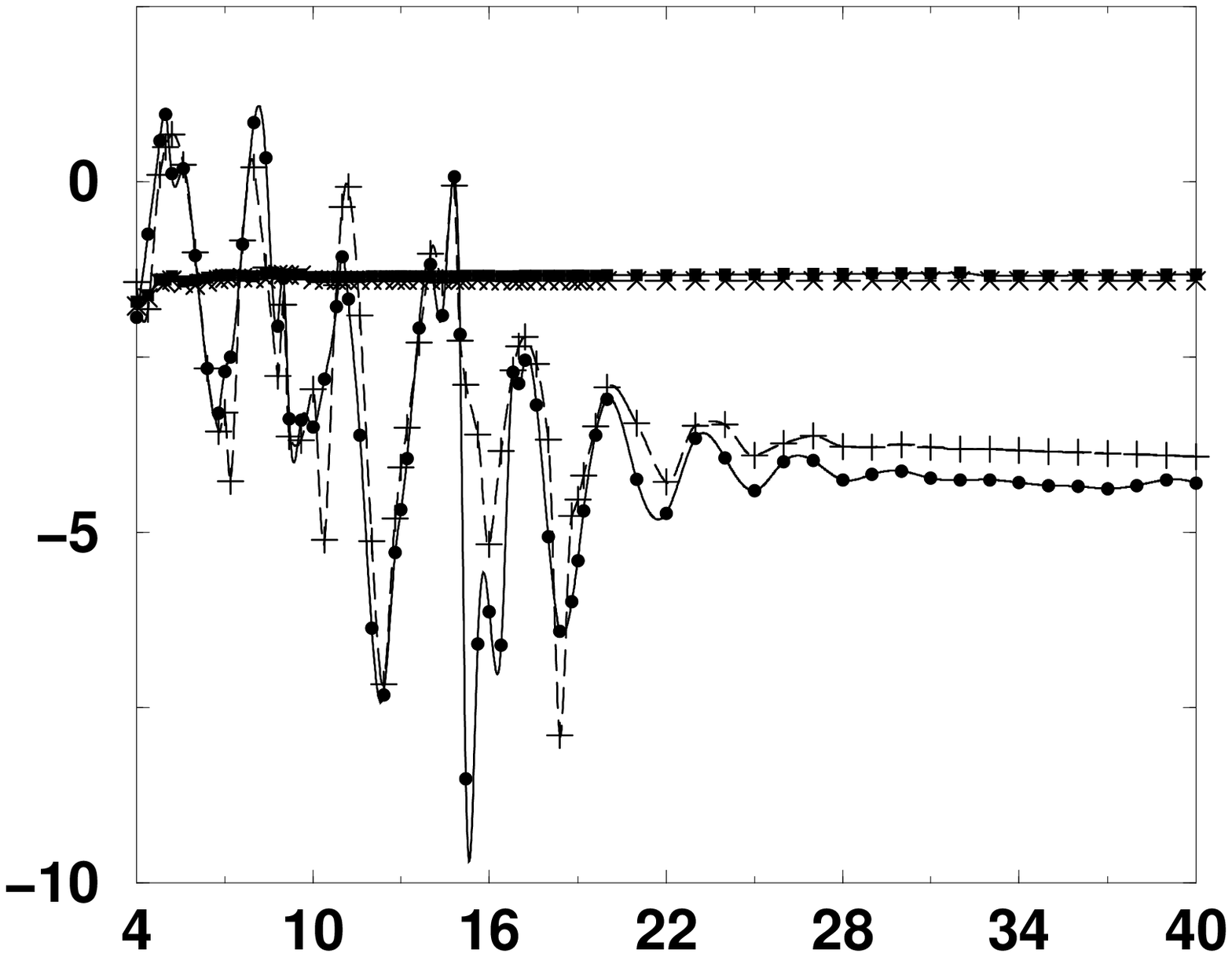}
\vspace*{5mm}
\hspace*{3mm}\includegraphics[width=6cm]{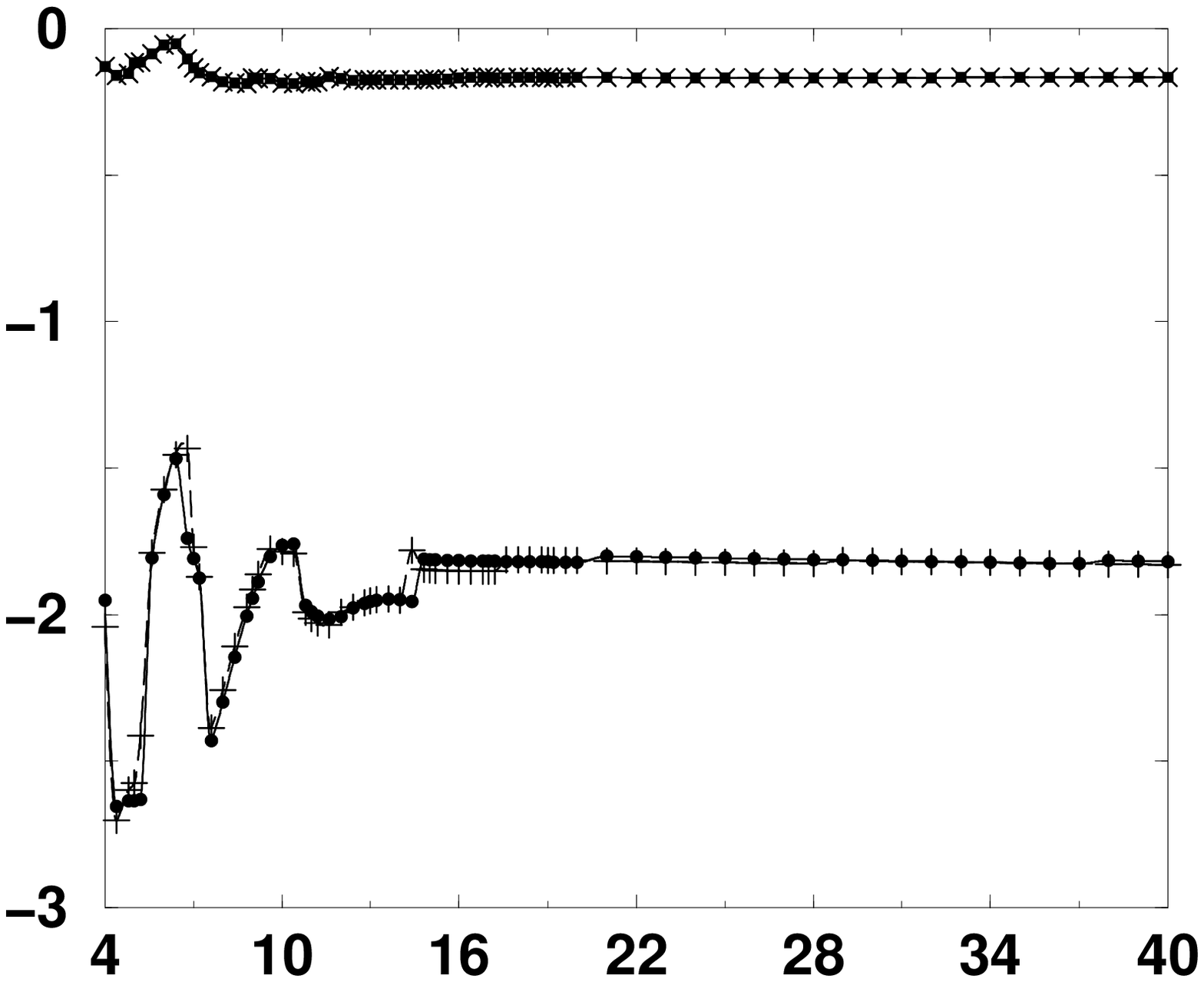}
\caption{ \CCae }
\end{figure}
}

The surface free energy as a function of cell width is shown in
Fig. \ref{fig:cw1}. In all cases the it reaches a
constant value for large cell widths (which is the surface tension). 
For large cell widths the walls
become isolated from each other with a layer of bulk
fluid intervening between the surface regions of fluid. At smaller
cell widths there is no bulk fluid layer so the surface layers
interact directly, leading to variation in the surface free energy with cell
width. This behaviour is generally oscillatory where the
distance between peaks is of the order of the molecular
length (2-3 $\sigma$).

Shown in Fig. \ref{fig:cw1}a is the surface free energy as a function of
cell width calculated using the numerical DCF. As can be seen the
surface free for the nematic fluid is of larger magntude than that
of the isotropic fluid and shows much stronger variation with cell
width. For the isotropic fluid it reaches a constant
value for cell widths greater than about 16 $\sigma$. In contrast in 
the nematic fluid it only approaches a constant value
for cell widths greater than about 25 $\sigma$. This results from the
larger range of the DCF in the nematic phase compared to in the
isotropic phase.

Figure \ref{fig:cw1}b shows the surface free energy as a function of cell
width calculated using the Parsons-Lee DCF. As for the numerical DCF
the surface free energy in the nematic phase is of greater magnitude and
shows stronger variation with cell width than in the isotropic phase.
In both the nematic and isotropic phases the it is
constant for $L$ greater than about 18 $\sigma$. This is in contrast
to the behaviour of the surface tension for the numerical DCF.

\section{Summary}

In this paper the structure of a fluid of soft ellipsoids near a soft
wall is determined using density functional theory calculations. The
calculated density and order parameters have been compared to
simulation results for the same system. Within the DFT calculations
the excess free energy was obtained using both a numerical direct
correlation function and a well known approximation to the DCF. Two 
different {\it Ans\"{a}tze} were used for the
excess free energy, giving very similar results to each other. 

There is qualitative agreement between the profiles found using the DFT
method and those from simulation. The agreement is better in the
isotropic phase, than for the nematic. In the isotropic case, the DFT
calculations give profiles that appear to be less structured than
those from simulation. This situation is reversed in the nematic phase.

For the isotropic phase, the Parsons-Lee DCF gave profiles similar to
those from the numerical DCF. However, in the nematic phase the
Parsons-Lee DCF gave profiles that were much less structured than
those from the numerical DCF and simulation.

The variation in the surface free energy as a function of cell width 
has also been examined. For larger cell
widths it is a constant, whereas for small cell widths
it reflects strong oscillatory interactions between
the two surfaces.

In summary, the simple density functionals used in this paper
which are based on a second order density expansion of 
the excess free energy, already give profiles which are
in qualitative agreement with simulations. We still
observe quantitative deviations, especially in the nematic
phase. Therefore it will be interesting to compare these
results with profiles obtained from more sophisticated
density functionals, {\em e. g.}, the Somoza-Tarazona
functional \cite{somoza1989a} or the recently proposed
modified Rosenfeld functional \cite{cinacchi2002a}.

\section*{Acknowledgements}

The authors wish to thank the German Science Foundation for funding. 
One of us (DLC) is grateful to Enrique Velasco for helpful advice. The
Monte Carlo program used was kindly porvided by Mark Wilson.


\bibliography{david1}
\bibliographystyle{unsrt}

\end{document}